\documentstyle[11pt,aaspp4]{article}  %preprint format
\def\sp{\sigma}
\def\rc{r_{core}}

\begin{document}

\title{NEW LIMITS ON A COSMOLOGICAL CONSTANT
FROM STATISTICS OF GRAVITATIONAL LENSING}

\author{Masashi Chiba}
\affil{National Astronomical Observatory, Mitaka, Tokyo 181, Japan}

\and

\author{Yuzuru Yoshii\altaffilmark{1}}
\affil{Institute of Astronomy, Faculty of Science, University of Tokyo, Mitaka,
Tokyo 181, Japan}
\altaffiltext{1}{Also at Research Center for the Early Universe,
School of Science, University of Tokyo, Bunkyo-ku, Tokyo 113, Japan}

%%%%%% Abstracts %%%%%%%%%%%%%%%%%%%%%%%%%%%%%%%%%%%%%%%%%%%%%%%%%%%%
\begin{abstract}
We present new limits on cosmological parameters from the statistics of
gravitational lensing, based on the recently revised knowledge of the
luminosity function and internal dynamics of E/S0 galaxies that are essential
in lensing high-redshift QSOs. We find that the lens models using updated
Schechter parameters for such galaxies, derived from the recent redshift
surveys combined with morphological classification, are found to give smaller
lensing probabilities than earlier calculated. Inconsistent adoption of these
parameters from a mixture of various galaxy surveys gives rise to systematic
biases in the results. We also show that less compact dwarf-type galaxies which
largely dominate the faint part of the Schechter-form luminosity function
contribute little to lensing probabilities, so that earlier lens models
overestimate incidents of small separation lenses. Applications of the
lens models to the existing lens surveys indicate that reproduction of
both the lensing probability of optical sources {\it and} the image separations
of optical and radio lenses is significantly improved in the revised lens
models. The likelihood analyses allow us to conclude that a flat universe with
$\Omega_0=0.3^{+0.2}_{-0.1}$ and $\Omega_0+\lambda_0=1$ is most preferable, and
a matter-dominated flat universe with $\lambda_0=0$
is ruled out at 98 \% confidence level. These new limits are unaffected by
inclusion of uncertainties in the lens properties.
\end{abstract}
%%%%%%%%%%%%%%%%%%%%%%%%%%%%%%%%%%%%%%%%%%%%%%%%%%%%%%%%%%%%%%%%%%%%%

%%--- luminosity function --- quasars: general --- structure}

%%% Sec.1 %%%
\section{INTRODUCTION}

Recent cosmological observations have increased various lines of evidence
that the matter density of the universe, $\Omega_0$, falls short of the closure
density required to halt its expansion (Ostriker \& Steinhardt 1995; Yoshii \&
Peterson 1995; Krauss 1997; Bahcall \& Fan 1998). These include
the anisotropies of the cosmic microwave background,
large-scale structure of galaxy distribution, number evolution of
clusters of galaxies, and large amount of gas in clusters combined with
the constraints on light-element nucleosynthesis.
Then, in order to reconcile with the flat geometry of a universe favored by
inflationary models, a cosmological constant $\Lambda$, or
$\lambda_0\equiv \Lambda c^2/3H_0^2$ where $H_0$ is the Hubble constant, has
been invoked to achieve the critical energy density, $\Omega_0+\lambda_0=1$.
In fact, the existence of a non-vanishing cosmological constant points to
a concordance of various observations and cosmological models with a
combination of $\Omega_0 \sim0.3$ and $\lambda_0\sim0.7$
(Ostriker \& Steinhardt 1995).
The case for a flat universe with a non-zero $\lambda_0$ appears to be
more likely than an open universe, even when the recent revisions of
the mean age of globular clusters and the current estimate of $H_0$ are
incorporated (Krauss 1997; Pont et al. 1998).
Most recently, the distance determinations of Type Ia supernovae at
cosmological distances (Perlmutter et al. 1998; Garnavich et al. 1998) revealed
further evidence for an accelerative expansion of a universe.
 
On the contrary, tight limits on a cosmological constant have been put from the
statistics of gravitational lensing, because the number of multiply imaged
QSOs found in lens surveys is sensitive to the value of $\lambda_0$
(Turner 1990; Carroll, Press \& Turner 1992). Systematic surveys to search
for lensed QSOs, such as the {\it Hubble Space Telescope} (HST) Snapshot
Lens Survey (Maoz et al. 1993),
have identified only a few lenses among hundreds of QSOs, and this small
lensing probability has been the central argument against a non-vanishing
cosmological constant (Maoz \& Rix 1993, hereafter referred to as MR).
In particular, Kochanek (1996) put severe limits $\lambda_0<0.66$ at 95 \%
confidence in flat cosmologies, using all of existing optical lens surveys
available then, together with the data of radio lenses to reproduce their
image-separation distributions. These constraints on $\lambda_0$ challenge
the proposed cosmic concordance required from other various observations.

Modeling the statistics of gravitational lensing however contains uncertainties
associated with the dynamical structure and number density of galaxies that
work as a lens. Among various galaxy types, elliptical (E) and lenticular (S0)
galaxies play an essential role in the models, because spiral and irregular
galaxies are known to make negligible contributions to lensing statistics
(MR; Kochanek 1996). Thus, attempts have been made to tighten the lens
properties from spectroscopic and photometric data of E/S0 galaxies
(Fukugita \& Turner 1991, hereafter FT; Breimer \& Sanders 1993; Franx 1993;
Kochanek 1994).
The internal dynamics of such galaxies is deduced from the velocity dispersions
and light profiles, whereas the luminosity function (LF) of E/S0 galaxies
provides the number density of galaxies responsible for lensing background
QSOs. Thus, the resultant constraints on $\lambda_0$ inevitably depend on how
well all of these lens properties are modeled.

Here we present new calculations of the lensing statistics in view of the
recently updated knowledge of the E/S0-type LFs and their internal dynamics.
We show that the models using the updated Schechter parameters for E/S0-type
LFs, which have been derived from the recent redshift surveys combined with
morphological classification, significantly revise the theoretical predictions
of both the lensing probabilities and image-separation distributions.
We also point out that earlier models considerably overestimate the lensing
effects of less compact dwarf-type galaxies, which dominate
the faint part of the Schechter-form LF. Our calculations using realistic internal
structure and dynamics of such galaxies in $\S$3 are found to diminish their
contribution to lensing statistics.
Our revised calculations applying to the data of the HST
Snapshot Lens Survey were reported in Chiba \& Yoshii (1997), showing that
a low-density, flat universe is not statistically inconsistent with
observations. This article is devoted to the further properties of our models
for the lensing statistics in more detail. In particular, in order to pursue
the most plausible set of cosmological parameters ($\Omega_0,\lambda_0$),
we use all of the optical lens surveys including the HST Snapshot Lens Survey.
The data of radio lenses are also taken into account to tighten the results of
the statistics (Kochanek 1996), by estimating the relative likelihood of their
observed image-separation distributions.
In $\S$2 we present the models for gravitational lensing and set all of
ingredients required in the lensing statistics. The effects of the E/S0-type
LFs on the lensing probability are discussed in $\S$3 to demonstrate how the
updated knowledge of the LFs modifies the results in a significant manner.
Then in $\S$4, we apply the current lens models to the existing lens data and
derive the new limits on cosmological parameters. Implications of our results
are discussed and conclusions are drawn in $\S$5.

%%% Sec.2 %%%
\section{GRAVITATIONAL LENSING OF QUASARS}
\subsection{{\it Lens model and lensing probability}}

In the presence of a lens along the light-ray path from a source to
an observer, the ray is deflected by an amount that depends on
the gravitational field of a lens.
We assume that the density profile of a lens is represented by that of an
isothermal sphere having a finite core,
\begin{equation}
\rho(r) = \frac{\sp^2}{2\pi G (r^2 + \rc^2)} \ ,
\end{equation}
where $\sp$ is the one-dimensional velocity dispersion and $\rc$ is the core
radius. The impact parameter of a deflected light in the lens plane, $b$,
is related to that of an undeflected light, $l$, through the equation
(Hinshaw \& Krauss 1987),
\begin{equation}
b + l = r_e \frac{(b^2+\rc^2)^{1/2} - \rc}{b} \ ,
\end{equation}
where $r_e$ is the critical impact parameter for multiple imaging in the case
of a singular isothermal sphere (SIS) as a lens. This is defined as $r_e=4\pi
(\sp/c)^2 D_{OL}D_{LS}/D_{OS}$, where $D_{OL}$ is the angular-diameter
distance between the observer and the lens, $D_{LS}$ between the lens and the
source, and $D_{OS}$ between the observer and the source. We use the filled
beam formulae to define these distances.

Equation (2) admits three solutions $b_i(l)$ with $i=1,2,3$
if $l$ is less than a critical impact parameter $l_0$ [eq.(A2)-(A4)], under
the condition that the ratio between $\rc$ and $r_e$, $\beta \equiv \rc/r_e$,
is smaller than 1/2. The image 1 lies on the same side from the lens as
the source, whereas the other two images (images 2 and 3) lie on the opposite
side of the lens from the source, where the
innermost image 3 is faintest among three images. If either $l > l_0$ or
$\beta > 1/2$ is fulfilled, there is only one image. We summarize the
characteristic properties of these lensed images in the Appendix (see also
Hinshaw \& Krauss 1987).
Each image is then magnified by a factor $A_i \equiv |b_i db_i/l dl|$
compared to an unlensed image, and the total magnification factor $A$ is
defined as $A \equiv A_1+A_2+A_3$. As demonstrated in the Appendix, the
presence of a core leads to larger $A$ than the SIS case, while the cross
section for lensing is reduced, $\pi l_0^2 < \pi r_e^2$.
The image separation angle $\Delta\theta$ for these multiple images is
well characterized by that between outer two images (images 1 and 2) for
$l = 0$, namely, $\Delta\theta = 2 (1-\beta)^{1/2} r_e / D_{OL}$.

Statistics of gravitational lensing depend on the number density distribution
of lens galaxies. This is deduced from the LF of galaxies
$\phi_g(L_g)$ with luminosity $L_g$. To compare with the earlier standard
models of lensing statistics, we adopt the assumption that the comoving number
density of galaxies is independent of lens redshift $z_L$.
The results of the Canada-France Redshift survey by Lilly et al. (1995)
support this assumption (see also Totani \& Yoshii 1998).

Then, the probability that a QSO with redshift $z_S$ and luminosity $L_Q$ is
lensed is evaluated from the fraction of the QSOs that are amplified to this
luminosity by lensing, among the total number of QSOs.
Using the LF of QSOs defined as $\phi_Q(L_Q)$, this is expressed as,
%the former is proportional to $\phi_Q(L_Q/A)d(L_Q/A)$, while the latter
%$\phi_Q(L_Q)dL_Q$. We thus obtain,
\begin{equation}
p(L_Q,z_S) = \int^{z_S}_0 dz_L (1+z_L)^3 \frac{cdt}{dz_L}
 \int^\infty_0 dL_g \phi_g(L_g) \int^{l_0(L_g,z_L)}_0 dl 2\pi l S
  \frac{\phi_Q(L_Q/A,z_S)/A}{\phi_Q(L_Q,z_S)} \ ,
\end{equation}
where $S$ denotes the selection function which depends on the image separation
and magnitude difference between the primary and secondary images, and on the
seeing in a lens survey (e.g. Maoz \& Rix 1993).
To perform the integration over $l$, it is convenient to avoid the
divergence of image magnification when a QSO lies at $l=0$ and $l_0$.
For this purpose, we integrate it over the position of image 2, $b_2$, instead
of the source position $l$, using the Jacobian $|A_2|$ which relates these two
positions (Blandford \& Narayan 1986). This yields
$\int dl 2\pi l \rightarrow \int db_2 2\pi b_2/|A_2|$ over the range of
$b_{-} \le b_2 \le b_{+}$, where $b_{+}$ and $b_{-}$ are the critical positions
of image 2 relevant to $l = 0$ and $l_0$, respectively.

In order to relate luminosities of galaxies $L_g$ to their structural
parameters ($\sp, \rc$), we first assume the scaling relation between $L_g$
and $\sp$, $L_g = L_g^\ast (\sp / \sp^\ast)^\gamma$,  where $\sp^\ast$
is the one-dimensional velocity dispersion at $L_g=L_g^\ast$. This is in
analogy to the Faber-Jackson relation for E and S0
galaxies, and we adopt $\gamma=4$ for these galaxies that effectively work as
a lens. As discussed in Kochanek (1996) and Chiba \& Yoshii (1997),
we will not introduce the correction to $\sp^\ast$ for the dark-matter
velocity dispersion by a factor of $(3/2)^{1/2}$, because detailed 
dynamical modeling of early-type galaxies by Breimer \& Sanders (1993),
Franx (1993), and Kochanek (1994) has invalidated this correction.
If this correction is employed, the models will overestimate the lensing
probability by a factor of $2.25$.
Second, in contrast to the simple treatment of $\rc=const$ in Chiba \&
Yoshii (1997), we take into account the observed increase of core radii
with increasing luminosities of E/S0 galaxies, by the relation
$L_g = L_g^\ast (\rc / \rc^\ast)^{1/\eta}$, where $\eta=1.2$ (FT; Kochanek
1996; Faber et al. 1997). From these scaling relations, the parameter
$\beta \equiv \rc/r_e$ is expressed as
$\beta = (\rc^\ast/r_e^\ast) (L_g/L_g^\ast)^{\eta-1/2}$,
thereby increasing with increasing $L_g$.

In conventional lens models, the LF of lens galaxies $\phi_g$ is assumed to
hold the Schechter form,
\begin{equation}
\phi_g(L_g) dL_g = \phi_g^\ast (L_g/L_g^\ast)^\alpha e^{-(L_g/L_g^\ast)}
 dL_g/L_g^\ast \ ,
\end{equation}
where $\phi_g^\ast$, $\alpha$, and $L_g^\ast$ are the normalization, the index
of faint-end slope, and the characteristic luminosity, respectively.
In this case, the lensing probability $p$ is proportional to a non-dimensional
factor $F^\ast$, defined as
\begin{equation}
F^\ast \equiv 16\pi^3 \left( \frac{\sp^\ast}{c} \right)^4 
       \left( \frac{c}{H_0} \right)^3 \phi_g^\ast
       \Gamma(\alpha + 4/\gamma + 1) \ ,
\end{equation}
(Turner, Ostriker \& Gott 1984; FT).

%%%%
\subsection{{\it LF of QSOs and magnification bias}}

Equation (3) indicates that the lensing probability depends on the shape of
the QSO LF, $\phi_Q (L_Q)$. As a standard case, we use the $B$-band LF adopted
by Wallington \& Narayan (1993) (hereafter WN) on the basis of
the LF determination for $z_S \le 3$ (Boyle et al. 1988) and that
for $z_S > 3$ (Schmidt et al. 1992; Warren et al. 1992; Irwin et al. 1992).
Written as a function of absolute magnitude, $M_Q$, the QSO LF
$\Phi_Q \equiv |\partial L_Q/\partial M_Q| \phi_Q$ is given in the smoothed
two power-law form
\begin{equation}
\Phi_Q(M_Q,z_S) \propto \frac{1}{
     10^{0.4(M_Q-M_z)(\beta_1+1)} + 10^{0.4(M_Q-M_z)(\beta_2+1)}  } \ ,
\end{equation}
where $M_z$ is the magnitude of the break in the LF. For $z_S < 3$, this is
given as
\begin{equation}
M_z = -2.5 k_L \log(1+z_S) + M_0 \ ,
\end{equation}
with $k_L=3.15$ and $M_0=-22.42 B$ mag for $h=0.5$
($h\equiv H_0/100$ km s$^{-1}$ Mpc$^{-1}$), $\Omega_0=1$, and $\lambda_0=0$.
For $z_S > 3$, the evolution of the break magnitude $M_z$ is corrected by
$-0.54(z_S-3)$ to accord with the high-redshift QSO surveys (Wallington \&
Narayan 1993). The parameters $\beta_1=-1.44$ and $\beta_2=-3.79$ denote the
faint and bright end slopes, respectively.

The effect of the so-called magnification bias such that lensed QSOs are
over-represented in magnitude-limited sample due to image amplification
is taken into account through the ratio $\phi_Q(L_Q/A,z_S)/A / \phi_Q(L_Q,z_S)$
in the integrand of eq.(3). In contrast to the SIS case,
we note that in the presence of a finite core, the integration over $l$ (or
over $b_2$) that includes this ratio cannot be taken outside of
other integrations
over $z_L$ and $L_g$ due to the dependence of $A$ on $\beta(z_L,L_g)$ (see also
Hamana et al. 1997 for more detail). To demonstrate the amount of the biasing
effect, designated as $B(L_Q,z_S)$, we use the following definition
\begin{equation}
B(L_Q,z_S) = \frac{p}{p(A \to 1)} \ .
\end{equation}
In the SIS case, this is reduced to
$B=\int_2^\infty dA (8/A^3) \phi_Q(L_Q/A,z_S)/A / \phi_Q(L_Q,z_S)$.

We calculate the bias $B(M_Q,z_S)$ as a function of $M_Q$ in the $B$ band,
using a set of fiducial lens parameters $\gamma=4$, $\eta=1.2$, and
$\sp^\ast=220$ km s$^{-1}$.
Figure 1(a) compares the SIS case (solid lines) with the case having a finite
core $\rc^\ast=0.1h^{-1}$ kpc for $\alpha=-1$ (dotted lines) and $\alpha=0$
(dashed lines).
It is evident that the bias is made large in the presence of
a core, so that the effect of smaller cross section for multiple imaging than
in the SIS case is reduced (Kochanek 1996; Hamana et al. 1997). This is
attributed to the larger image amplification $A$ in the presence of a core.
It is also noted that the shallower faint-end slopes $\alpha$ in the LFs of
galaxies yield the larger bias, which thereby reduces the effect of less
numbers of galaxies in such LFs on the lensing probability. This arises from
the fact that
the LFs having the shallower $\alpha$ weight toward larger $L_g$ galaxies, thus
larger $\beta$, while the magnification bias increases with increasing $\beta$
(Hamana et al. 1997).
Figure 1(b) shows the effects of changing QSO LFs on the bias.
In comparison with WN's QSO LF,
we also consider the broken power-law QSO LFs adopted by MR ($\beta_1=-1.2,
\beta_2=-3.6, k_L=3.5, M_0=-21.75$ mag), and FT ($\beta_1=-1.7, \beta_2=-3.15,
{\rm break\ in\ apparent\ magnitude}=19.15$ mag).
When compared to WN's LF, FT's LF yields the significant
reduction of the bias at the bright end due to its much shallower slope
$\beta_2$, whereas MR's LF yields roughly the same bias although it is
slightly smaller at high redshifts $z_S$.
This suggests that the lensing probability based on either FT's or MR's LF
will be smaller than that based on WN's LF, as was also noted in Kochanek
(1996).

%%% Sec.3 %%%
\section{E/S0-TYPE LUMINOSITY FUNCTIONS AND LENSING PROBABILITIES}

\subsection{{\it Effects of updated Schechter parameters}}

Many of earlier lensing statistics use the Schechter-form LF of all-type
galaxies in Efstathiou et al. (1988) scaled by the Postman \& Geller (1984)
fraction (31\%) of E/S0 galaxies, because these early-type galaxies are
essential in lensing statistics (FT; MR). The corresponding E/S0-type LF is
hereafter referred to as EEP's LF and its Schechter parameters
($M^\ast,\alpha,\phi_g^\ast$) are tabulated in Table 1, where $M^\ast$ is the
characteristic magnitude derived from $L_g^\ast$.
Although this procedure of counting
the number density of E/S0 galaxies has widely been adopted for lensing
statistics, the different sampling and calibration between these
observational studies prevent us from knowing the detailed shape of the
E/S0-type LF in the sample of Efstathiou et al. (1988). Specifically,
the faint-end slope $\alpha$ and characteristic magnitude $M^\ast$
for only E/S0 galaxies
are not necessarily the same as those for all-type galaxies. Thus, it is not
surprising that the lensing statistics based on EEP's LF holds some
unavoidable systematics.

Comparing with the above indirect method, the type-dependent LF for only E/S0
galaxies has been obtained on the basis of the recent redshift surveys of
field galaxies combined with morphological classification.
We consider the two LFs derived from the Stromlo-APM survey (Loveday et al.
1992, LPEM) and the CfA survey (Marzke et al. 1994, MGHC), in addition to
EEP's LF. The Schechter parameters for only E/S0 galaxies are taken from these
papers and tabulated in Table 1.
Comparing with EEP's LF, LPEM's LF is characterized by the shallow
slope $\alpha$ at the faint end and the small normalization $\phi_g^\ast$,
whereas for MGHC's LF, $M^\ast$ is about 1 mag fainter and $\phi_g^\ast$ is
about twice larger than the respective values of EEP's LF.
These revised values of the Schechter parameters will yield the different
values of $\sp^\ast$ and $F^\ast$ from those based on EEP.

We note that the high normalization $\phi_g^\ast$
of MGHC's LF may be explained in terms of local density
inhomogeneities present in the CfA survey volume, while the Stromlo-APM survey
samples a volume $\sim30$ times larger than that of the CfA survey and so
provides the reliable space density of galaxies. However, the reason for the
faint value of $M^\ast$ in MGHC's LF is not well understood. There are some
discussion that this is due to the fainter Zwicky magnitudes used in the CfA
survey than the $b_J$ magnitudes in the Stromlo-APM survey, by an amount of
$0.2-0.5$ mag (see e.g. Huchra 1976; Auman et al. 1986;
Bothun \& Cornell 1990, about detailed discussions of the Zwicky magnitudes).
If we transform $M^\ast$ in MGHC by adding a constant offset,
then the resultant value of $M^\ast$ is shifted toward the values in EEP and
LPEM. To see how this offset changes the lensing probability, we also consider
the case of $M^\ast \to M^\ast+0.5$ mag in MGHC, hereafter referred to as MGHC2
(Table 1).
We note that there are also some suggestions that the Zwicky magnitudes depend
on surface brightness of galaxies, in a manner that the luminosities of
compact galaxies such as E's tend to be overestimated (Auman et al. 1986;
de Vaucouleurs et al. 1991). If this is the case, $M^\ast$ for E galaxies
measured from the Zwicky magnitudes may be brighter than from other magnitude
systems. As discussed in EEP,
the transforms between the magnitude systems also contain color terms, complex
isophotal corrections, etc., so that the procedure of transforming $M^\ast$
from one system into another by adding a constant offset is a gross
oversimplification. Thus, in view of these uncertainties, the results based on
MGHC2 need caution.

Most recently, the Las Campanas Redshift Survey (LCRS) provided the expanded
galaxy sample in the largest survey volume (Lin et al. 1996). The LF of all
galaxy sample (after transformation from the Kron-Cousins $R$ band system used
in LCRS to the $b_J$ band) is in good agreement with that from the Stromlo-APM
survey. Although the type-dependent LFs are not yet available,
the LFs based on the [O II] $\lambda$3727 emission feature (Lin et al. 1996)
or on the more detailed spectral classification scheme (Bromley et al. 1997)
show that the faint-end slopes $\alpha$ become progressively shallower from
late to early-type galaxies. This dependence of $\alpha$ on galaxy type
is in good agreement with LPEM's result.
Thus, an argument that the shallow $\alpha$ in LPEM's E is due to
failure of identifying faint ellipticals (MGHC) may not be supported.
Here for comparison, we also take into account LCRS's LF based on
the galaxies that show no [O II] $\lambda$3727 emission (Table 1).
We shift LCRS's $M^\ast$ by 1.1 mag to match the mean rest-frame color
$<b_J-R>_0=1.1$ of LCRS galaxies (Lin et al. 1996). It is worthwhile to
note that the derived $M^\ast$ is slightly fainter than those in other LFs.
However, the results also
need great caution because of ambiguous morphological classification.

In order to derive the characteristic velocity dispersion $\sp^\ast$ from the
observed $M^\ast$, we use the Faber-Jackson relation
determined by de~Vaucouleurs \& Olson (1982) for early-type galaxies,
\begin{eqnarray}
-M^\ast + 5\log h &=& 19.37 + 10 (\log \sp_E^\ast - 2.3) 
   \qquad \mbox{\rm for E} \nonumber \\
-M^\ast + 5\log h &=& 19.75 + 10 (\log \sp_{S0}^\ast - 2.3)
   \qquad \mbox{\rm for S0} \ .
\end{eqnarray}
Then, for the fractional values ($f_E,f_{S0}$) of (E, S0) galaxies in the
number density, respectively, we evaluate $\sp^\ast$ for E/S0 galaxies by means
of ${\sp^\ast}^4=(f_E {\sp_E^\ast}^4 + f_{S0} {\sp_{S0}^\ast}^4)/(f_E+f_{S0})$.
This is because the lensing probability is proportional to the fourth power of
$\sp^\ast$ in the definition of $F^\ast$ (eq.5).
We adopt the Postman \& Geller (1984) fraction (E:12\%, S0:19\%)
to calculate $\sp^\ast$ for the LFs of EEP, LPEM, and LCRS, while for the LFs
of MGHC and MGHC2 we use Marzke et al. (1994)'s Table 1.
The values of $\sp^\ast$ and $F^\ast$ derived from these new LFs appear to be
systematically smaller than those from EEP's LF (see Table 1)
due to the above mentioned difference in the Schechter parameters.
Therefore, the total lensing probability based on the updated Schechter
parameters for E/S0 galaxies will be reduced.

To highlight these effects of updating the Schechter parameters,
we calculate the lensing probability of a QSO with $M_Q=-25.5+5\log h$ mag
and $z_S=2$, for the standard lens parameters of $\gamma=4$,
$\rc^\ast=0.1h^{-1}$ kpc, and $\eta=1.2$. Figure 2 shows the image-separation
distributions $p(\Delta\theta)$ for the cases of EEP (solid line),
LPEM (dotted line), MGHC (short-dashed line), MGHC2 (long-dashed line), and
LCRS (thick solid line). The image-separation
distribution for the case of EEP differs significantly from those for LPEM and
MGHC. LPEM's LF having shallower faint-end slope $\alpha$ and smaller
normalization $\phi^\ast_g$ shifts the distribution to somewhat large image
separations with smaller overall amplitude. MGHC's LF having fainter
$M^\ast$ considerably suppresses the distribution at larger separation and
makes it peaked at smaller separation. In either case, the total
lensing probability is decreased when compared to the result from EEP's LF.
This clearly indicates that the predicted lensing probability is
sensitive to the adopted Schechter parameters for the LFs of E/S0 galaxies.
The figure also suggests that the case of MGHC2 produces a similar
$p(\Delta\theta)$ to that of EEP, whereas the case of LCRS leads to the
skewed $p(\Delta\theta)$ towards larger image separations with smaller overall
amplitude. Considering great ambiguities in the E/S0-type LFs inferred from
MGHC2 and LCRS as discussed above, we will not consider these
cases in what follows.

Updated LFs of LPEM and MGHC hold separate combinations of Schechter
parameters and yield different lensing probabilities. However, Kochanek (1996)
contrived the LF in a mixed manner by adopting $M^\ast$ from the all-type
LF by EEP, $\alpha$ from the all-type LF by LPEM and MGHC, and $\phi_g^\ast$
from the all-type LF by LPEM after scaled with MGHC's fraction of
E/S0 galaxies. In his lens model, the characteristic velocity dispersion
$\sp^\ast$ for E/S0 was taken from the work of Kochanek (1994) that showed
$\sp^\ast=225$ km s$^{-1}$ using EEP's $M^\ast$. To compare with other
cases, we also consider the LF adopted by Kochanek (1996), hereafter referred
to as K96's LF, in the statistics discussed in $\S$4, and the associated
Schechter and lens parameters are given in the last line of Table 1.
We remark that in each galaxy survey, the Schechter parameters have been
determined in a highly correlated manner, so that Kochanek's method of
combining the results from various surveys lacks consistency
and thus causes artificial systematics. In particular, while the
number density $\phi_g^\ast$ of E/S0 galaxies was averaged between LPEM and
MGHC, adoption of the bright $M^\ast=-19.9$ mag and steep $\alpha=-1$,
irrespective of the reported fainter $M^\ast$ in LPEM and the shallower
$\alpha$ in MGHC for E/S0 galaxies than each other, leads to a bias in favor
of large lensing probabilities. The best way of avoiding these artificial
systematics is to use the Schechter parameters as reported in the original
references.

%%%
\subsection{{\it Effects of deviation from the Schechter form}}

Earlier models for lensing statistics adopted the Schechter form for the
E/S0-type LF. Specifically, the index of faint-end slope $\alpha$ is usually
$\simeq -1.0$ as in EEP, thereby implying the existence of numerous faint E/S0
galaxies that work as a lens. However, the faint part of the E/S0-type LF is
still uncertain because the limited resolution hinders reliable type
classification for faint galaxies. Moreover, distance determinations for
{\it field} galaxies from their observed redshifts may be uncertain to some
extent because of the peculiar galactic motions in the local field.
On the other hand, the problem of accurate distance determinations is avoided
by using nearby clusters of galaxies which offer a homogeneous sample of
member galaxies at the same distance, observed down to faint magnitudes.
Sandage, Binggeli \& Tammann (1985) and Ferguson \& Sandage (1991) have
indicated that while the faint part of the LFs for all galaxies is dominated by
diffuse and less luminous dwarf-type galaxies, the LFs for E/S0 galaxies alone
also show a sharp decline towards faint magnitudes, similarly towards bright
magnitudes. This bounded form of LFs at both bright and faint ends was first
recognized by Hubble (1936). For field E/S0 galaxies, there is also an
indication of such a decline towards faint magnitudes (Binggeli, Sandage \&
Tammann 1988; Driver et al. 1994; Driver et al. 1995).

It has also become clear that early-type dwarf galaxies below
$M = -17 \sim -18 B$ mag which populate at the faint end of the LF show
the different internal structure and dynamics from those of bright ellipticals
(e.g. Ferguson \& Binggeli 1994). The radial brightness profile of such dwarf
galaxies is well represented by the exponential form
$I(r) \propto \exp(-r/r_d)$, where $r_d$ is the scale length. This is
reminiscent of the profile of spiral disks and is in contrast with the more
centrally concentrated $r^{1/4}$ law in bright ellipticals. This brightness
profile suggests a large core radius
$\rc \simeq 1$ kpc (e.g. Kormendy 1988; Ferguson \& Binggeli 1994),
so that the parameter $\beta \equiv \rc/r_e$ introduced in the current lens
model may be larger than 1/2 in all lens redshifts.
Also, the luminosity-velocity relation follows $L \propto \sp^{2.5}$
(Held et al. 1992) that deviates from the $L \propto \sp^4$ law of the bright
E/S0 galaxies. Thus, the faint galaxies have smaller internal velocity
dispersions than that expected from the $L \propto \sp^4$ law.
These diffuse properties of faint early-type galaxies are well predicted from
the models of formation of dwarf galaxies invoking galactic winds and
subsequent adiabatic expansion (Dekel \& Silk 1986; Yoshii \& Arimoto 1987).
Consequently, because these faint early-type galaxies have large $\rc$ and
small $\sp$, their contribution to the lensing statistics is negligible.

To demonstrate these properties of dwarf galaxies in lensing calculations,
we construct the model LFs of early-type galaxies which consist of both E/S0
and dE/dS0 types. The parameters for the LFs used are tabulated in Table 2 and
the shape of the LFs is graphically displayed in Fig.3. The Schechter
parameters for E/S0-type galaxies roughly
accord with those derived from existing galaxy surveys, except
that the LF at the faint magnitude is reduced by multiplying a factor
$\exp(-10^{0.4(M-M^{cut})})$ with $M^{cut}=-17$ mag in order to match the
observed shape of E/S0-type LFs (Driver et al. 1995). The Schechter parameters
for dE/dS0-type galaxies are taken from the work of Phillips and Driver
(1995), by assuming $M^\ast(dE/dS0)=M^\ast(E/S0)+3$ mag,
$\phi_g^\ast(dE/dS0)=\phi_g^\ast(E/S0)$, and a very steep slope $\alpha=-1.5$.
As is evident from Fig. 3, the total LF (thin solid line) derived from the
E/S0-type (dashed line) and dE/dS0-type LFs (dotted line) resembles a single
Schechter form. The parameters by which this total LF is fitted to
a single Schechter function (thick solid line) are also given in Table 2.

Using these LFs, we calculate the lensing probability of a QSO with
$M_Q=-25.5+5\log h$ mag and $z_S=2$. For E/S0 galaxies, the lens parameters
are taken as $\gamma=4$, $\rc^\ast=0.1h^{-1}$ kpc, and $\eta=1.2$, whereas
for dE/dS0 galaxies, we assume $\gamma=2.5$,
$\rc=0.2h^{-1}$ kpc $\simeq constant$ ($\eta=0$). 
In order to evaluate $\sp^\ast$ from $M^\ast$ for E/S0 galaxies, we use eq.(9)
and the Postman \& Geller (1984) fraction for these galaxy types.
For dE/dS0 galaxies, we use the data of IC794 (Bender \& Nieto 1990;
Held et al. 1992) to obtain the following approximate relation that obeys
the scaling $L_g/L_g^\ast = (\sp/\sp^\ast)^{2.5}$,
\begin{equation}
-M^\ast + 5\log h = 16.00 + 6.25 (\log \sp^\ast - 1.72)
   \qquad \mbox{\rm for dE/dS0} \ .
\end{equation}

Figure 4(a) shows the lensing probability at $z_L=0.5$ as a function of
galaxy magnitudes $M$ in the $B$ band. It follows that dE/dS0 galaxies
contribute very little to the lensing probability; if we adopt a more realistic
core radius of $\rc \simeq 1$ kpc (e.g. Kormendy 1988; Ferguson
\& Binggeli 1994), the lensing probability is much more reduced than presented
with $\rc \simeq 0.2$ kpc.
Figure 4(b) shows the image separation distribution. It is evident from these
figures that use of a simple Schechter form for E/S0-type LF, as has been
adopted in previous lensing models, considerably overestimates the lensing
probability at the faint magnitudes and therefore at the small image
separations.

In order to consider this trivial effect of dwarf-type galaxies on
lensing probabilities at the faint part of the LFs of EEP, LPEM, and MGHC,
we introduce a magnitude cutoff of $\exp(-10^{0.4(M-M^{cut})})$ in these LFs
and examine the effect of changing $M^{cut}$ over its possible range.
Figure 5(a) shows the image-separation distributions $p(\Delta\theta)$ for
$M^{cut}=+\infty$ (no cutoff), $-16$, and $-17$ mag, and Fig. 5(b) shows
the ratios of the total lensing probability $p$ with cutoff relative to that
without cutoff as a function of $M^{cut}$. It is clear that the suppression of
faint, less-massive galaxies from the originally Schechter-form LF reduces
the incidents of small-separation lensing. We also find that the result from
LPEM is very insensitive to $M^{cut}$, because the shallow faint-end slope
$\alpha$ of LPEM's LF implies the small number of faint E/S0 galaxies
irrespective of introducing the magnitude cutoff.

%%% Sec.4 %%%
\section{APPLICATION TO THE EXISTING LENS SURVEYS}

\subsection{{\it QSO sample and selection functions}}

We apply the current lens model to the sample of 867 unduplicated QSOs at
$z_S>1$ (as in Kochanek 1996), which are taken from the optical lens surveys
such as the CFHT Survey (Crampton et al. 1992; Yee et al. 1993), the ESO
Key-Programme Survey (Surdej et al. 1993), the HST Snapshot Survey (Maoz et al.
1993), the NOT Survey (Jaunsen et al. 1995), and the FKS Survey (Kochanek
et al. 1995). The $B$-band absolute magnitudes of these QSOs are
evaluated from the $V$-band apparent magnitudes tabulated in each survey,
a power-law QSO spectrum of $\propto \nu^{-0.5}$ for the $K$-corrections,
and a typical $B-V=0.2$ mag in the QSO rest frame as adopted by
previous workers. We note that $V$ magnitudes of some numbers of QSOs are based
on the catalogue of V\'eron-Cetty and V\'eron (1996) which provides not
always accurate magnitudes. In this combined sample, only six QSOs are lensed
(1208$+$1011, 1413$+$117, 1009$-$0252, 1115$+$080, 0142$-$100, and
0957$+$561). We do not use 0957$+$561 having a large image separation
$\Delta\theta=6.''1$, because the lensed image of this QSO is affected by
an intervening cluster of galaxies.

When we apply the lens model to these lens surveys, the selection functions
$S$ (see eq.3) inherent in the surveys for detecting multiple images give
the additional constraints on the range of the impact parameter for lensing and
thus affect the estimate of lensing probabilities. For each QSO, we evaluate
the selection function which depends on the seeing in the relevant survey and
the magnitude difference between the primary and secondary images.
When the QSOs are observed in both the HST and other grand-based
surveys, we adopt the HST results because of the high resolution
of detecting separate images down to $\Delta\theta = 0''.1$.

In addition to the optical sample, we consider the radio lenses,
as in Kochanek (1996). While the incomplete information on the
redshift and $B$-band luminosities of the sample in the radio surveys prevents
us from calculating the absolute lensing probability in this sample, it is
possible to use the observed image separations of the radio lenses in order to
obtain how likely these data are reproduced in the current lens models.
We adopt the ten radio lenses listed in Kochanek (1996)
(CLASS 1608$+$656, MG 1131$+$045,
MG 0414$+$0534, MG 1654$+$1346, B 1938$+$666, MG 1549$+$3047,
CLASS 1600$+$434, B 1422$+$231, MG 0751$+$2716, and B 0218$+$357).
For the purpose of comparing his results with ours, we follow his procedure to
assign redshifts and magnitudes for the lenses when these data are not
available. 

We then calculate the sum of the lensing probabilities $p_i$ for the optical
QSOs to obtain the expected number of lensed QSOs $n=\sum p_i$ and
the image-separation distribution $n(\Delta\theta)=\sum p_i(\Delta\theta)$.
The former is restricted to the lenses with $\Delta\theta \le 4''$ in order to
avoid ambiguities associated with unusually large separations. We also
calculate the likelihood function $L$
for the models to reproduce $N_U$ unlensed optical QSOs, $N_L$ lensed optical
QSOs, and $N_R$ radio lenses having image separations $p(\Delta\theta)$ (see
Kochanek 1996),
\begin{equation}
\ln L = - \sum_{i=1}^{N_U}p_i + \sum_{j=1}^{N_L}\ln p_j(\Delta\theta_j)
    + \sum_{k=1}^{N_R}\ln \left( \frac{p_k(\Delta\theta_k)}{p_k} \right) \ ,
\end{equation}
where only the relative likelihood is computed for radio lenses to achieve
the observed image separations.

\subsection{{\it Results}}

Figure 6(a) shows the predicted number of optical lenses
$n(\Delta\theta \le 4'')$ against $\Omega_0$ for the flat universe of
$\Omega_0+\lambda_0=1$.
We take a standard parameter set of $\rc^\ast=0.1h^{-1}$ kpc, $\gamma=4$,
$\eta=1.2$, and WN's LF for QSOs. Different curves correspond to the
results from the LFs of EEP (solid line), LPEM (dotted line),
MGHC (dashed line), and K96 (thick solid line). For EEP, LPEM, and MGHC,
three cases of $M^{cut}=-16.5$, $-17.0$, and $-17.5$ mag are shown from
the lower to upper curves, whereas the magnitude cutoff is not used for K96.
Comparing the predicted number of lenses with
the observed five lenses among the current QSO sample, it is obvious that
different LFs predict the different best values of $\Omega_0$,
as $\Omega_0\sim0.2$ for MGHC, $0.3$ for LPEM, and $0.5$ for EEP.
This effect of adopting different LFs overwhelms the effect of changing
$M^{cut}$ significantly. It is remarkable that use of the updated LFs for E/S0
galaxies yields a small $\Omega_0$ (or large $\lambda_0$) in contrast to the
earlier strong conclusion in favor of $\Omega_0=1$ (MR; K96).

In order to elucidate the most likely case, we show in Fig.6(b) the
image-separation distribution $n(\Delta\theta)$ for each LF with the value of
$\Omega_0$ that reproduces the observed number of lenses.
The histogram denotes the image-separation distribution of the five optical
lenses identified in the current sample, and the distribution of the ten radio
lenses is shown by marking an asterisk at each image separation.
Comparing the predicted and observed distribution $n(\Delta\theta)$
including radio lenses, the model using LPEM's LF appears to reproduce
the data better than the models using other LFs.
Specifically, MGHC's LF yields too many small separations and at the same
time falls short of the observed number of both optical and radio lenses
at $2''<\Delta\theta<3''$. This large difference between the predicted and
observed separation distributions holds also for the LFs of both EEP and K96.

Following these results, we determine the best combination of LF and
cosmological parameters that maximizes the likelihood $L$ defined in eq. (11)
to simultaneously reproduce the number of lenses $n(\Delta\theta \le 4'')$
and the separation distribution $n(\Delta\theta)$.
Figure 7 shows the likelihood results when using the LFs of LPEM
(dotted lines), EEP (solid lines), and K96 (thick solid line),
where three lines for LPEM and EEP correspond to the cases with
$M^{cut}=-16.5$, $-17$, and $-17.5$ mag. All likelihood values are normalized
by its maximum $L_{max}$ which is derived using LPEM's LF with
$M^{cut}=-17$ mag. Note that
the case for MGHC's LF yields quite a small $L$ ($\ln L/L_{max}<-19$) and
is not shown in the figure. We have confirmed Kochanek (1996)'s result that an
$\Omega_0\simeq1$ universe gives the highest likelihood
($\ln L/L_{max} \sim -2.8$) as long as his LF is used.
However, EEP's LF with $M^{cut} \simeq -17$ mag also provides the
similar likelihood at a lower $\Omega_0$.
It is evident from these cases that LPEM's LF yields much larger
likelihood over a broad range of $\Omega_0$. The peak of the likelihood
is located at $\Omega_0 \simeq 0.3$ in flat cosmologies, and this appears
to be very insensitive to the value of $M^{cut}$. If we take $\Omega_0=0.3$
as the most likely case, a universe with $\Omega_0=1$ is ruled out at
98 \% confidence level. For an open universe with $\lambda_0=0$, the likelihood
monotonically increases with decreasing $\Omega_0$ and $\ln L/L_{max}$ is
as small as $-1.5$ at $\Omega_0=0.1$ with LPEM's LF.

\subsection{{\it Uncertainties}}

The above analyses support that the observations are best explained by the
lens model using the updated E/S0-type LF of LPEM, among the LFs adopted in the
current work.
It should be kept in mind that MGHC's LF contains some uncertainties related
to the Zwicky magnitude system used in the CfA survey. We then examine
how the likelihood based on LPEM's LF is changed when the inherent
model parameters are varied, and the result is tabulated in Table 3.
Decreasing a core radius does not affect the likely range
of $\Omega_0$ significantly, because the effect of magnification bias is
simultaneously decreased, as discussed in $\S$2.2. Artificial increase of
$\sp^\ast$ irrespective of the value derived from the Faber-Jackson relation
(eq. 9) tends to favor a larger $\Omega_0$, but the likelihood value turns out
to decrease at the same time, as discussed below. Intriguingly, the likelihood
result for LPEM's LF is essentially unchanged even if no cutoff is employed,
because of the shallow faint-end slope $\alpha$.
Use of MR's QSO LF slightly decreases the likely value of $\Omega_0$, but the
change is modest.

The lensing probability is proportional to ${\sp^\ast}^4$, thereby
the statistics based on the observed number of lensed QSOs is sensitive to
what value is assigned to $\sp^\ast$. We have estimated $\sp^\ast$ from the
Faber-Jackson relation (eq. 9) using the observed value of $M^\ast$ and
the possible fraction of E and S0 types. However, while the large fraction
of S0-type galaxies for a given fraction of both E and S0 types implies
smaller $\sp^\ast$, as suggested from eq.(9), there are large uncertainties in
distinguishing between E and S0 types. Also, the self-consistent modeling of
internal dynamics in E/S0 galaxies suggests that $\sp$ increases slowly with
increasing $\rc$ (Kochanek 1996). In view of these uncertainties in assigning
the value of $\sp^\ast$, we in turn compute the likelihood for various values
of $\sp^\ast$ in flat cosmologies, using LPEM's LF.

Figure 8 shows the likelihood in the two dimensional parameter
space $(\sp^\ast, \Omega_0)$ for flat cosmologies. Contours are shown
at 68 \% (1 $\sigma$), 90 \%, 95.4 \% (2 $\sigma$), and 99 \% confidence
levels for one degree of freedom in the likelihood ratio.
It follows that within the 1 $\sigma$ confidence interval for $\sp^\ast$,
which includes $\sp^\ast\simeq 205$ km s$^{-1}$ derived from the Faber-Jackson
relation (see table 1), a low-density universe with $\Omega_0\sim0.3$ is
favored. If we further increase $\sp^\ast$, the peak of the likelihood is
located at a higher $\Omega_0$ but the overall likelihood turns to decrease.
This is because the models with such large values of $\sp^\ast$
yield larger average image separations than observed.

Therefore, even if uncertainties in the lens model are taken into account,
a low-density, flat universe with a large $\lambda_0$ is statistically
consistent with the observations when we use the updated LF for E/S0 galaxies.
In sharp contrast to the previous models of lensing statistics
that have supported a high-density universe with $\Omega_0=1$,
we conclude that a universe with $\Omega_0=0.3^{+0.2}_{-0.1}$ and
$\Omega_0+\lambda_0=1$ casts the best case to explain the results of the
observed lens surveys.

%%% Sec.5 %%%
\section{DISCUSSION AND CONCLUSIONS}

We have presented new calculations of gravitational lensing statistics in view
of the recently revised knowledge of the internal dynamics and number density
of early-type galaxies. Main revised points in the lens models are summarized
as follows. 
(1) The factor of $(3/2)^{1/2}$ correction for $\sp$ is not used following
the detailed dynamical modeling of early-type galaxies (Breimer \& Sanders
1993; Franx 1993; Kochanek 1994). (2) The type-specific Schechter parameters
for E/S0 galaxies are adopted from the recent redshift surveys of galaxies
combined with morphological classification. (3) The faint
part of the LF of E/S0 galaxies, which is dominated by diffuse dwarf-type
galaxies, is found to be unimportant for lensing statistics. All of these
revisions point to smaller lensing probabilities than earlier predicted.
Applications of the lens models to the existing lens surveys suggest that
both the total number of optical lenses {\it and} the image separations of
optical and radio lenses are best reproduced in a low-density, flat
universe. As the best set of cosmological parameters we arrive at
$\Omega_0=0.3^{+0.2}_{-0.1}$ with $\Omega_0+\lambda_0=1$. A flat universe
with $\Omega_0=1$ is ruled out at 98 \% confidence. We also find that
$\lambda_0 < 0.9$ at 94 \% confidence in flat cosmologies, and that
an open universe with $\lambda_0=0$ is less likely than a flat universe with
$\lambda_0 \ne 0$.

Our conclusion supporting a large $\lambda_0$ is virtually in contrast to
that of Kochanek (1996) who strongly argued against it using the same QSO
sample as adopted here. Main difference between his result and ours originates
from the points (2) and (3) stated above. In particular, Kochanek gathered the
Schechter parameters of E/S0 galaxies from various LF determinations by EEP,
MGHC, and LPEM. He has adopted $M^\ast=-19.9$ mag from the all-type LF by
EEP, $\alpha=-1$ from the all-type LF by LPEM and MGHC, and $\phi_g^\ast=
6.1\times10^{-3}$ from the scaling of LPEM's $\phi_g^\ast=14.0\times10^{-3}$
for all types by MGHC's E/S0 fraction. However, since the Schechter
parameters have been determined by means of a highly correlated fitting in each
galaxy survey, a simple mixture of these parameters taken from various
references lacks consistency. Specifically, his use of brighter $M^\ast$
and steeper $\alpha$ than those of E/S0 galaxies reported in the original
references (see Table 1) leads to a bias in favor of a systematically
large lensing probability. A further effect against a large $\lambda_0$ seen
in his model as well as other earlier models lies in the counting of large
numbers of less compact dwarf-type galaxies as lens objects which dominate
the faint part of the LF.

Models for gravitational lensing must explain not only the observed probability
of lensing but also the relative probability of showing a specified image
separation. The average image separation increases with increasing $\sp^\ast$,
thereby the lens data are expected to constrain $\sp^\ast$ irrespective of
the value derived from $M^\ast$ found in each galaxy survey and the
Faber-Jackson relation. However, our exercise of using the trivial effect
of faint galaxies in lensing, which is also consistent with shallower faint-end
slope $\alpha$ for earlier types seen in LPEM and LCRS (Lin et al. 1996;
Bromley et al. 1997), implies a larger average image
separation even when $\sp^\ast$ is fixed. As a result, the estimate of
$\sp^\ast$ which accords with the image-separation distributions depends on
the faint part of the LF for E/S0 galaxies adopted. In this respect, it is
notable that the lens model using LPEM's $\alpha=0.2$ which was not explored
by Kochanek (1996) provides much better fitting to the observed image
separations for optical and radio lenses, even if his preferred value of
$\sp^\ast \simeq 220$ km s$^{-1}$ is used in our analysis.

Intriguingly, our 1 $\sigma$-confidence limit $0.2\le\Omega_0\le0.5$ with
$\Omega_0+\lambda_0=1$ is in good accord with the recent result of
Im, Griffiths, \& Ratnatunga (1997) who analyzed the redshifts of known lensed
QSOs combined with the observed image separations and lens properties.
It is worthwhile to note that their method is immune to the uncertainties
associated with the determinations of the LF. This may suggest that lensing
statistics using the realistic form for the LF, as adopted in the present work,
yield the similar results. However, before concluding so definitely, we require
a sample of more high-$z$ QSOs in light of gravitational lensing,
together with more definite knowledge of the internal dynamics and number
density of lensing galaxies over a wide range of redshifts. Significant
increase and improvement of the information on lensing phenomena will then
allow us to answer whether there is a concordance of a flat,
$\lambda_0$-dominated universe with observations.

%%%%%%%%%%%%%%%%%%%%%%%%%%%%%%%%%%%%%%%%%%%%%%%%%%%%%%%%%
\acknowledgments

The authors are grateful to T.~Futamase and B.~A.~Peterson for useful
discussions during the course of the present work.
This work has been supported in part by the Grand-in-Aid
for Scientific Research (09640328) and COE Research (07CE2002)
of the Ministry of Education, Science, and Culture in Japan.

%%%%%%%%%%%%%%%%%%%%%%%%%%%%%%%%%%%%%%%%%%%%%%%%%%%%%%%%%
\appendix
\section{The solutions to the lens equation}

From the lens equation (2), the critical impact parameter $l_0$ for multiple
imaging is written as
\begin{equation}
l_0^2 = r_e^2 [ (1 + 5\beta - \frac{1}{2}\beta^2) - \frac{1}{2} \beta^{1/2}
        (\beta + 4)^{3/2} ] \ ,
\end{equation}
provided $\beta \equiv \rc/r_e < 1/2$.
After some algebra, the three solutions to eq.(2) are given by
\begin{eqnarray}
b_1 &=& -\frac{2}{3} l + 2 p^{1/2} \cos \frac{\phi}{3} \\
b_2 &=& -\frac{2}{3} l + 2 p^{1/2} \cos (\frac{\phi}{3}+\frac{2\pi}{3}) \\
b_3 &=& -\frac{2}{3} l + 2 p^{1/2} \cos (\frac{\phi}{3}-\frac{2\pi}{3}) \ ,
\end{eqnarray}
where
\begin{eqnarray}
\tan \phi &=& \frac{\sqrt{4p^3-q^2}}{-q} \\
p &=& \frac{1}{3} (r_e^2 - 2r_e \rc + \frac{l^2}{3}) \\
q &=& \frac{2l}{3} (r_e^2 + r_e \rc - \frac{l^2}{9}) \ .
\end{eqnarray}

Figure 9(a) shows the total amplification factor $A \equiv A_1+A_2+A_3$
for two representative cases of $\beta=0.1$ and 0.25, compared to the
SIS case. It follows that the presence of a finite core leads to diverging
$A$ at the caustics $l=0$ and $l_0$, and hence results in the value of $A$
which is larger than the SIS case.
Figure 9(b) shows the cross section $\sigma = \pi l_0^2$
for multiple imaging (thick solid line), and the separation angles
$\Delta\theta$ between two outer images (image 1 and 2) for $l=0$
(solid line), $0.7l_0/r_e$ (dotted line), and $0.9l_0/r_e$ (dashed line),
respectively. It is clear that the presence of a finite core reduces the
cross section for lensing, and that $\Delta\theta$ is almost independent
of the impact parameter $l$, thereby being well approximated as
$\sqrt{1-2\beta}$ at $l=0$.

%%%%%%%%%%%%%%%%%%%%%%%%%%%%%%%%%%%%%%%%%%%5

\clearpage
%%%%%%%%%%%%%%%%%%%%%%%%%%%%%%%%%%%%%%%%%%%%%%%%%%%%%%%%%

\clearpage
%%%%%%%%%%%%%%%%%%%%%%%%%%%%%%%%%%%%%%%%%%%%%%%%%%%%%%%
%%% table 1 %%%%%%%%%%%%%%%%%%%%%%%%%%%%%%%%%%%%%%%%%%%
%%%%%%%%%%%%%%%%%%%%%%%%%%%%%%%%%%%%%%%%%%%%%%%%%%%%%%%
\begin{center}
TABLE 1  \\
\medskip

E/S0 L{\sc uminosity} F{\sc unctions}$^a$
{\sc and} L{\sc ens} P{\sc arameters}

\medskip
\begin{tabular}{lccccc}
\hline\hline 
%%%%%%%%%%%%%%%%%%%%%%%%%%%%%%%%%%%%%%%%%%%%%%%%%%%%%%%
%%%%%%%%%%%%%%%%%%%%%%%%%%%%%%%%%%%%%%%%%%%%%%%%%%%%%%%
Survey & $M^\ast$  & $\alpha$ & $\phi_g^\ast$ & $\sp^\ast$    & $F^\ast$  \\
       & ($B$ mag) &          & (Mpc$^{-3}$)  & (km s$^{-1}$) &           \\
\hline
%%%%%%%%%%%%%%%%%%%%%%%%%%%%%%%%%%%%%%%%%%%%%%%%%%%%%%%
EEP$^b$...........& $-19.90$ & $-1.10$ & $4.8\times 10^{-3}$ & 214.5 & 0.018 \\
LPEM$^c$......... & $-19.71$ & $+0.20$ & $3.2\times 10^{-3}$ & 205.3 & 0.010 \\
MGHC$^d$........  & $-18.99$ & $-0.90$ & $9.1\times 10^{-3}$ & 166.0 & 0.011 \\
MGHC2$^e$......   & $-19.49$ & $-0.90$ & $9.1\times 10^{-3}$ & 186.2 & 0.017 \\
LCRS$^f$......... & $-19.12$ & $-0.27$ & $1.1\times 10^{-2}$ & 179.2 & 0.017 \\
K96.............. & $-19.90$ & $-1.00$ & $6.1\times 10^{-3}$ & 225.0 & 0.026 \\
%%%%%%%%%%%%%%%%%%%%%%%%%%%%%%%%%%%%%%%%%%%%%%%%%%%%%%%
\hline
%%%%%%%%%%%%%%%%%%%%%%%%%%%%%%%%%%%%%%%%%%%%%%%%%%%%%%%
%%%%%%%%%%%%%%%%%%%%%%%%%%%%%%%%%%%%%%%%%%%%%%%%%%%%%%%
\end{tabular}
\begin{flushleft}
\hspace*{2.7cm}$^a$ In the $B$ band and $h=1$. \\
\hspace*{2.7cm}$^b$ $\phi_g^\ast$ for all galaxy types is scaled by the E/S0
fraction (31\%) from \\
\hspace*{2.4cm}Postman \& Geller (1984). \\
\hspace*{2.7cm}$^c$ $\phi_g^\ast$ is taken from the fit by Driver, Windhorst,
\& Griffiths (1995). \\
\hspace*{2.4cm}$\sp^\ast$ is calculated from eq.(9) weighted by the Postman \&
Geller fraction of \\
\hspace*{2.4cm}E (12\%) and S0 (19\%) galaxies. \\
\hspace*{2.7cm}$^d$ $M^\ast$ is an average of $-19.23B$ mag (E) and
 $-18.74B$ mag (S0), and \\
\hspace*{2.4cm}$\alpha$ is an average of $-0.85$ (E) and $-0.94$ (S0).
$\sp^\ast$ is calculated from eq.(9) \\
\hspace*{2.4cm}weighted by MGHC's fractions of E and S0 galaxies. \\
\hspace*{2.7cm}$^e$ $M^\ast$ is transformed from that of MGHC by 0.5 mag. \\
\hspace*{2.7cm}$^f$ For galaxies with no [O {\sc II}] $\lambda3727$ emission
lines and with the mean color \\
\hspace*{2.4cm}$<b_J-R>_0=1.1$ mag. $\sp^\ast$ is from eq. (9) with the Postman
\& Geller fraction. \\
\hspace*{2.7cm}R{\sc eferences}.--EEP, Efstathiou et al. (1988); LPEM, Loveday
et al. \\
\hspace*{2.4cm}(1992); MGHC, MGHC2, Marzke et al. (1994); LCRS, Lin et al. (1996); \\
\hspace*{2.4cm}K96, Kochanek (1996) \\
\end{flushleft}
\end{center}

\clearpage
%%%%%%%%%%%%%%%%%%%%%%%%%%%%%%%%%%%%%%%%%%%%%%%%%%%%%%%
%%% table 2 %%%%%%%%%%%%%%%%%%%%%%%%%%%%%%%%%%%%%%%%%%%
%%%%%%%%%%%%%%%%%%%%%%%%%%%%%%%%%%%%%%%%%%%%%%%%%%%%%%%
\begin{center}
TABLE 2  \\
\medskip

M{\sc odel} P{\sc arameters} {\sc for} LF{\sc s} {\sc of}
E/S0 {\sc and} dE/dS0 T{\sc ypes}$^a$

\medskip
\begin{tabular}{lcccc}
\hline\hline 
%%%%%%%%%%%%%%%%%%%%%%%%%%%%%%%%%%%%%%%%%%%%%%%%%%%%%%%
%%%%%%%%%%%%%%%%%%%%%%%%%%%%%%%%%%%%%%%%%%%%%%%%%%%%%%%
Galaxy Type & $M^\ast$  & $\alpha$ & $\phi_g^\ast$  & $M^{cut}$ \\
            & ($B$ mag) &          & (Mpc$^{-3}$)   & ($B$ mag) \\
\hline
%%%%%%%%%%%%%%%%%%%%%%%%%%%%%%%%%%%%%%%%%%%%%%%%%%%%%%%
E/S0$^b$................& $-19.7$ & $-1.0$ & $3.2\times 10^{-3}$ & $-17.0$ \\
dE/dS0.............& $-16.7$ & $-1.5$ & $3.2\times 10^{-3}$ & ...   \\
Both Types$^c$......& $-20.0$ & $-1.4$ & $1.6\times 10^{-3}$ & ...  \\
%%%%%%%%%%%%%%%%%%%%%%%%%%%%%%%%%%%%%%%%%%%%%%%%%%%%%%%
\hline
%%%%%%%%%%%%%%%%%%%%%%%%%%%%%%%%%%%%%%%%%%%%%%%%%%%%%%%
%%%%%%%%%%%%%%%%%%%%%%%%%%%%%%%%%%%%%%%%%%%%%%%%%%%%%%%
\end{tabular}
\begin{flushleft}
\hspace*{2.7cm}$^a$ In the $B$ band and $h=1$. \\
\hspace*{2.7cm}$^b$ The Schechter function is multiplied by a factor
$\exp(-10^{0.4(M-M^{cut})})$. \\
\hspace*{2.7cm}$^c$ Fit to a single Schechter form for the combined LFs of
both E/S0 \\
\hspace*{2.4cm}and dE/dS0 types. \\
\end{flushleft}
\end{center}

\clearpage
%%%%%%%%%%%%%%%%%%%%%%%%%%%%%%%%%%%%%%%%%%%%%%%%%%%%%%%
%%% table 3 %%%%%%%%%%%%%%%%%%%%%%%%%%%%%%%%%%%%%%%%%%%
%%%%%%%%%%%%%%%%%%%%%%%%%%%%%%%%%%%%%%%%%%%%%%%%%%%%%%%
\begin{center}
TABLE 3  \\
\medskip

R{\sc esults} {\sc of} M{\sc aximum} L{\sc ikelihood}
{\sc for} O{\sc ptical} {\sc and} R{\sc adio} L{\sc enses}

\medskip
\begin{tabular}{lccc}
\hline\hline 
%%%%%%%%%%%%%%%%%%%%%%%%%%%%%%%%%%%%%%%%%%%%%%%%%%%%%%%
%%%%%%%%%%%%%%%%%%%%%%%%%%%%%%%%%%%%%%%%%%%%%%%%%%%%%%%
Model & Maximum$^b$ & 68\%($1\sigma$) confidence
    & 90\% confidence   \\
\hline
%%%%%%%%%%%%%%%%%%%%%%%%%%%%%%%%%%%%%%%%%%%%%%%%%%%%%%%
Standard Case$^a$............. 
   & $\Omega_0\simeq 0.30$ & $0.17<\Omega_0<0.49$ & $0.12<\Omega_0<0.66$ \\
\hline\hline
%%%%%%%%%%%%%%%%%%%%%%%%%%%%%%%%%%%%%%%%%%%%%%%%%%%%%%%
%%%%%%%%%%%%%%%%%%%%%%%%%%%%%%%%%%%%%%%%%%%%%%%%%%%%%%%
Parameter Change &  &  &     \\
\hline
%%%%%%%%%%%%%%%%%%%%%%%%%%%%%%%%%%%%%%%%%%%%%%%%%%%%%%%
$\rc^\ast$ $\to$ $0.05h^{-1}$ kpc...... 
   & $\Omega_0\simeq 0.30$ & $0.19<\Omega_0<0.54$ & $0.14<\Omega_0<0.76$ \\
$\rc^\ast$ $\to$ $0.03h^{-1}$ kpc...... 
   & $\Omega_0\simeq 0.35$ & $0.21<\Omega_0<0.59$ & $0.15<\Omega_0<0.84$ \\
$\sp^\ast$ $\to$ 210 km s$^{-1}$..........
   & $\Omega_0\simeq 0.30$ & $0.19<\Omega_0<0.53$ & $0.14<\Omega_0<0.74$ \\
$\sp^\ast$ $\to$ 220 km s$^{-1}$..........
   & $\Omega_0\simeq 0.40$ & $0.25<\Omega_0<0.65$ & $0.20<\Omega_0<0.91$ \\
$M^{cut}$ $\to$ $+\infty$ mag.........
   & $\Omega_0\simeq 0.30$ & $0.18<\Omega_0<0.50$ & $0.13<\Omega_0<0.70$ \\
$\phi_Q$ $\to$ MR.....................
   & $\Omega_0\simeq 0.25$ & $0.13<\Omega_0<0.40$ & $0.05<\Omega_0<0.56$ \\
%%%%%%%%%%%%%%%%%%%%%%%%%%%%%%%%%%%%%%%%%%%%%%%%%%%%%%%
\hline
%%%%%%%%%%%%%%%%%%%%%%%%%%%%%%%%%%%%%%%%%%%%%%%%%%%%%%%
%%%%%%%%%%%%%%%%%%%%%%%%%%%%%%%%%%%%%%%%%%%%%%%%%%%%%%%
\end{tabular}
\begin{flushleft}
\hspace*{0.5cm}$^a$ Our standard choice of the parameters includes
no $(3/2)^{1/2}$ correction for $\sp^\ast$, \\
LPEM's $\phi_g$, WN's $\phi_Q$, $\rc^\ast=0.1h^{-1}$ kpc,
and $M^{cut}=-17B$ mag, for a flat universe  \\
with $\Omega_0+\lambda_0=1$.   \\
\hspace*{0.5cm}$^b$ The value of $\Omega_0$ at $L=L_{max}$ when
the binning interval for $\Omega_0$ is 0.05. \\ 
\hspace*{0.5cm}R{\sc eferences}.--LPEM, Loveday et al. (1992);
WN, Wallington \& Narayan (1993);   \\
MR, Maoz \& Rix (1993) \\
\end{flushleft}
\end{center}
%%%%%%%%%%%%%%%%%%%%%%%%%%%%%%%%%%%%%%%%%%%%%%%%%%%%%%%
%%%%%%%%%%%%%%%%%%%%%%%%%%%%%%%%%%%%%%%%%%%%%%%%%%%%%%%

\clearpage
%%%%%%%%%%%%%%%%%%%%%%%%%%%%%%%%%%%%%%%%%%%%%%%%%%%%%%%%%
%%%%%%%%%%%%%%%%%%%%%%%%%%%%%%%%%%%%%%%%%%%%%%%%%%%%%%%%%

\begin{figure}
\plotone{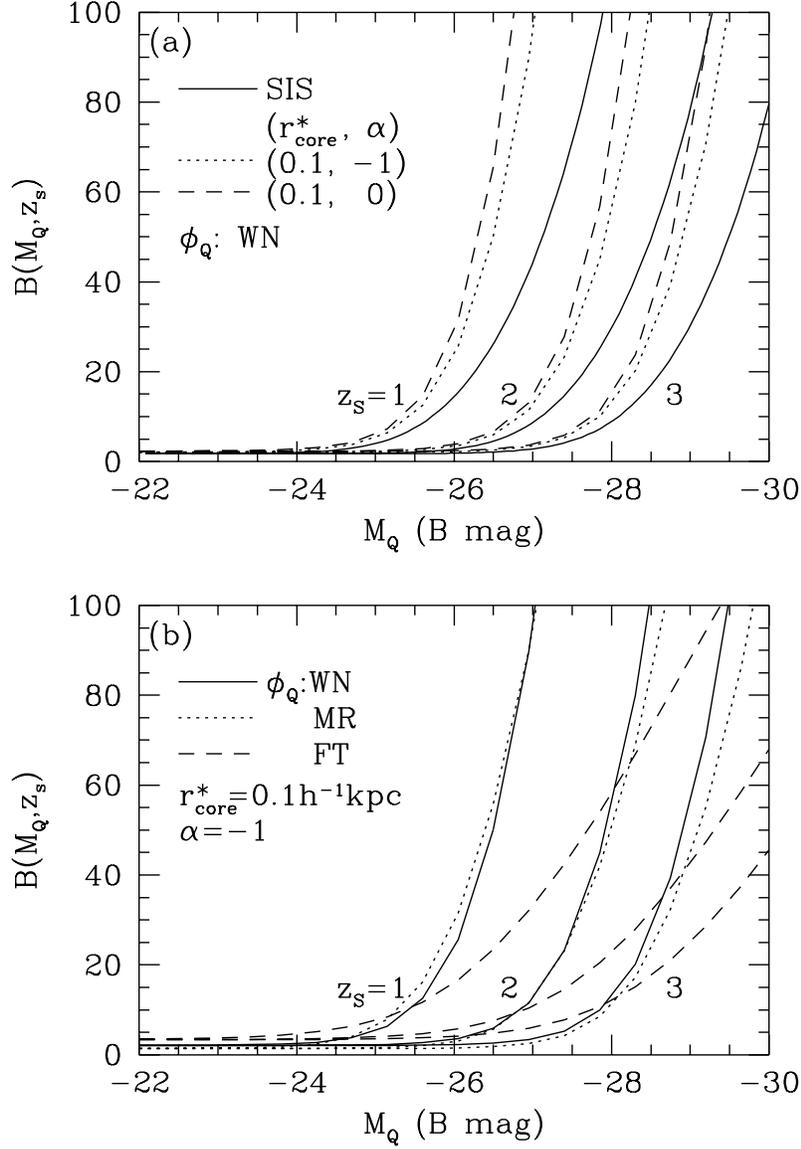}
\caption{
%\figcaption[fig1.ps]{
Magnification bias $B$ defined in eq.(8) as
a function of QSO magnitudes $M_Q$ for $h=0.5$. We adopt $\gamma=4$,
$\eta=1.2$, and $\sp^\ast=220$ km s$^{-1}$ for the lens parameters.
(a) Comparison between the SIS case (solid lines) and
the case with a finite core $\rc^\ast=0.1h^{-1}$ kpc for $\alpha=-1$
(dotted lines) and $\alpha=0$ (dashed lines). WN's QSO LF is utilized.
Note that presence of a core leads to large magnification bias
and that this effect becomes larger for shallower faint-end slopes
$\alpha$ of the LFs of galaxies.
(b) Effects of using different QSO LFs on bias (WN: solid lines,
MR: dotted lines, FT: dashed lines) for the case of $\rc^\ast=0.1h^{-1}$ kpc
and $\alpha=-1$. WN's QSO LF appears to give the largest bias.
}
\end{figure}

\clearpage
\begin{figure}
\plotone{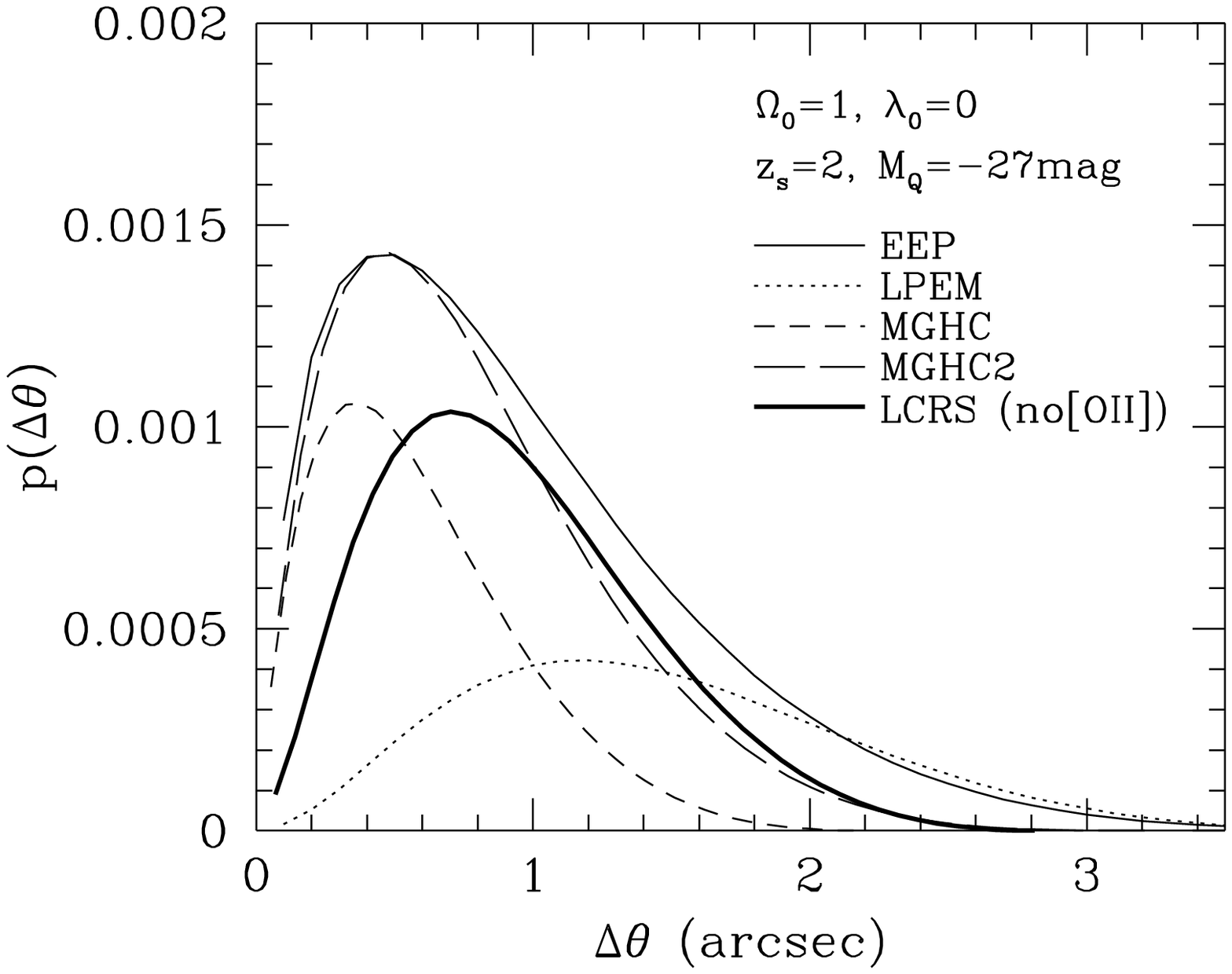}
\caption{
%\figcaption[fig2.ps]{
Lensing probability of a QSO with $M_Q=-27$ mag
(for $h=0.5$) and $z_S=2$, as a function of image separations
$\Delta\theta$. Various Schechter-form LFs of E/S0 galaxies, as tabulated
in Table 1, are examined, for a universe with $(\Omega_0,\lambda_0)=
(1,0)$.
}
\end{figure}

\clearpage
\begin{figure}
\plotone{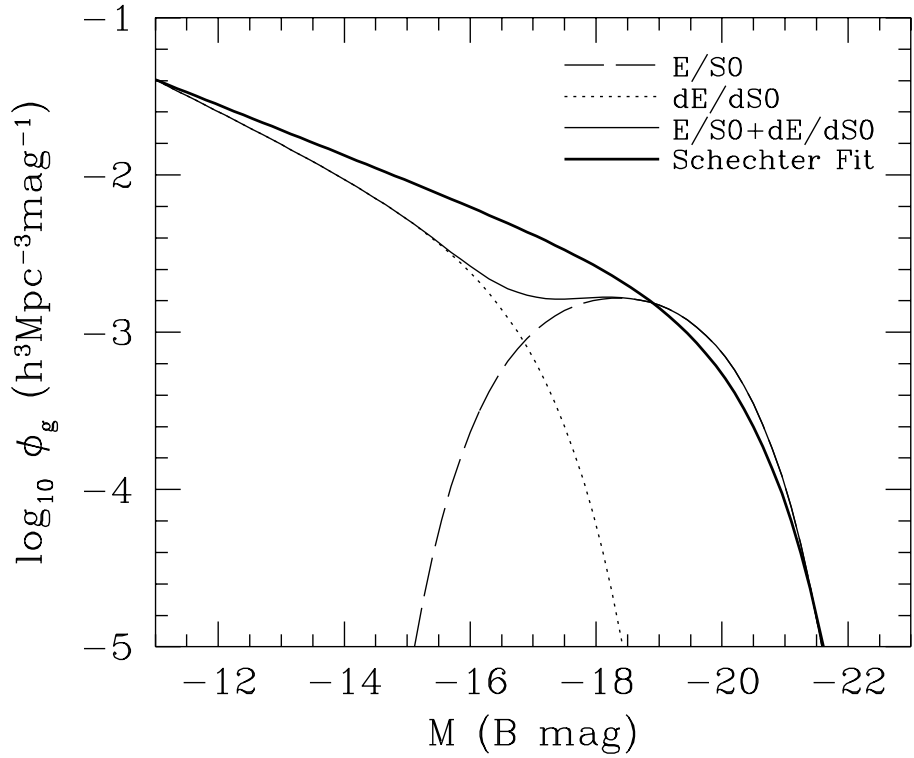}
\caption{
%\figcaption[fig3.ps]{
Model LFs of early-type galaxies as a function
of galaxy magnitudes $M$ (for $h=1$) are shown using the parameters given
in Table 2. Dashed and dotted lines denote
the LFs of E/S0 and dE/dS0 galaxies, respectively, and solid line
denotes the combined LF of these galaxies. Approximate fit of the combined
LF to a single Schechter form is shown by thick solid line. 
}
\end{figure}

\clearpage
\begin{figure}
\plotone{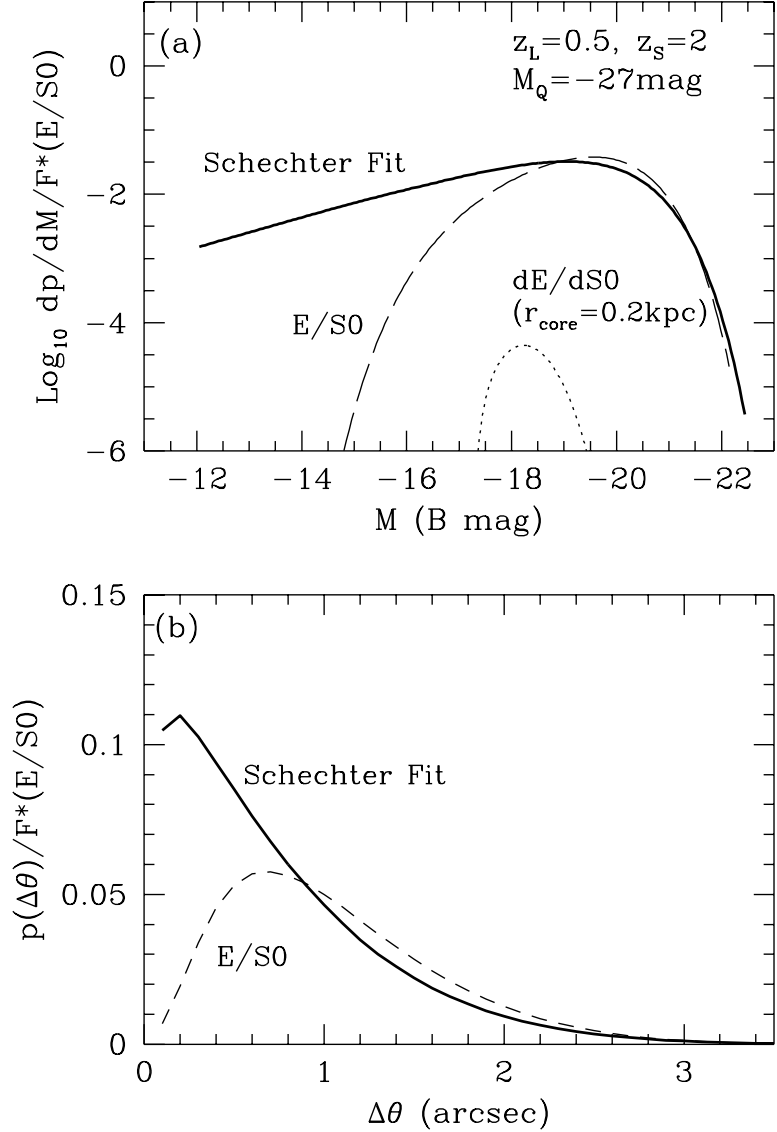}
\caption{
%\figcaption[fig4.ps]{
(a) Lensing probability at $z_L=0.5$ per unit
galaxy magnitude $dp/dM$ (normalized by $F^\ast$ for E/S0 galaxies),
when we use the model LFs shown in Fig.3.
Calculations are made for a QSO with $M_Q=-27$ mag
(for $h=0.5$) and $z_S=2$, and for $(\Omega_0,\lambda_0)=(1,0)$. 
See the text for the adopted lens parameters relevant to each galaxy type.
(b) Image-separation distribution $p(\Delta\theta)$ predicted from
the model LFs. The case for dE/dS0 galaxies is not shown because of
its giving negligible probability.
}
\end{figure}

\clearpage
\begin{figure}
\plotone{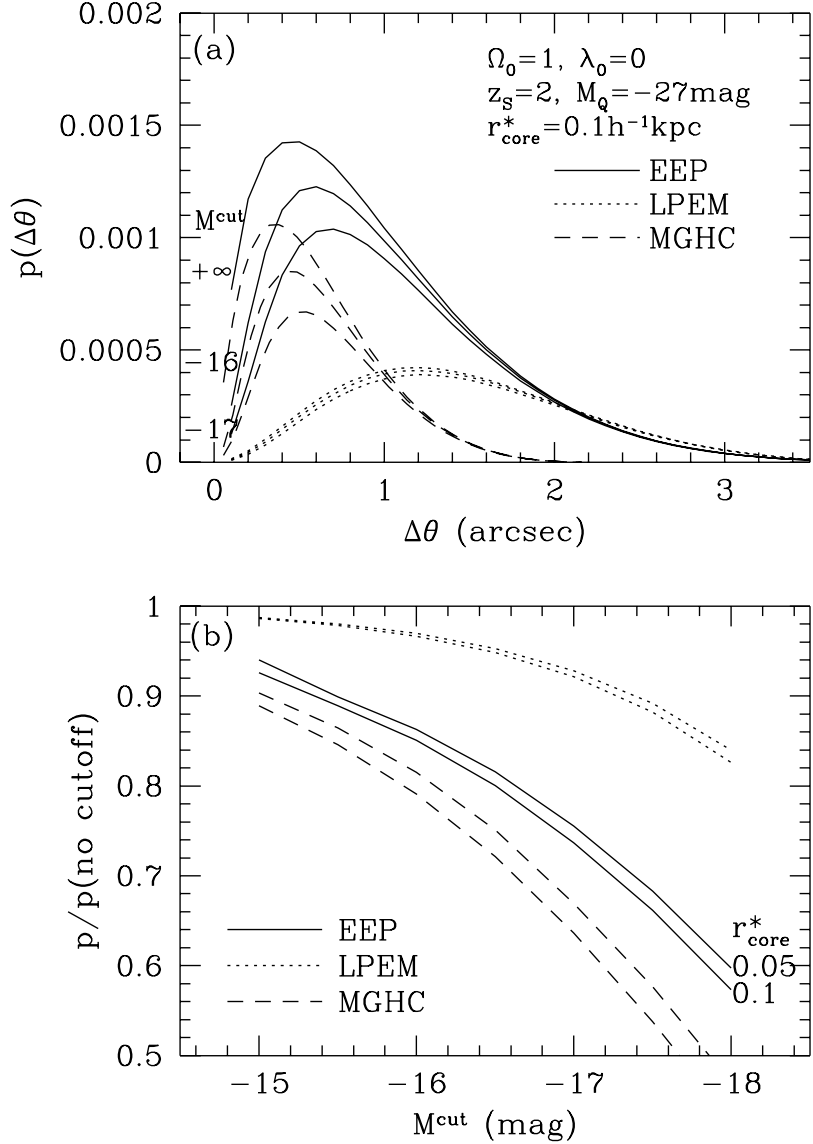}
\caption{
%\figcaption[fig5.ps]{
Effects of introducing a cutoff magnitude $M^{cut}$
for the LFs of EEP (solid lines), LPEM (dotted lines), and MGHC (dashed lines),
on the lensing probability. Calculations are made for a QSO with
$M_Q=-27$ mag (for $h=0.5$) and $z_S=2$, and for $(\Omega_0,\lambda_0)=(1,0)$.
(a) Image-separation distribution $p(\Delta\theta)$ for $M^{cut}=+\infty$
(no cutoff), $-16$, and $-17$ mag (for $h=1$), and
for $\rc^\ast=0.1h^{-1}$ kpc.
(b) Ratio of the total lensing probability $p$ with cutoff relative to that
without cutoff, as a function of $M^{cut}$. Two cases
for $\rc^\ast=0.05$ and $0.1h^{-1}$ kpc are shown.
}
\end{figure}

\clearpage
\begin{figure}
\plotone{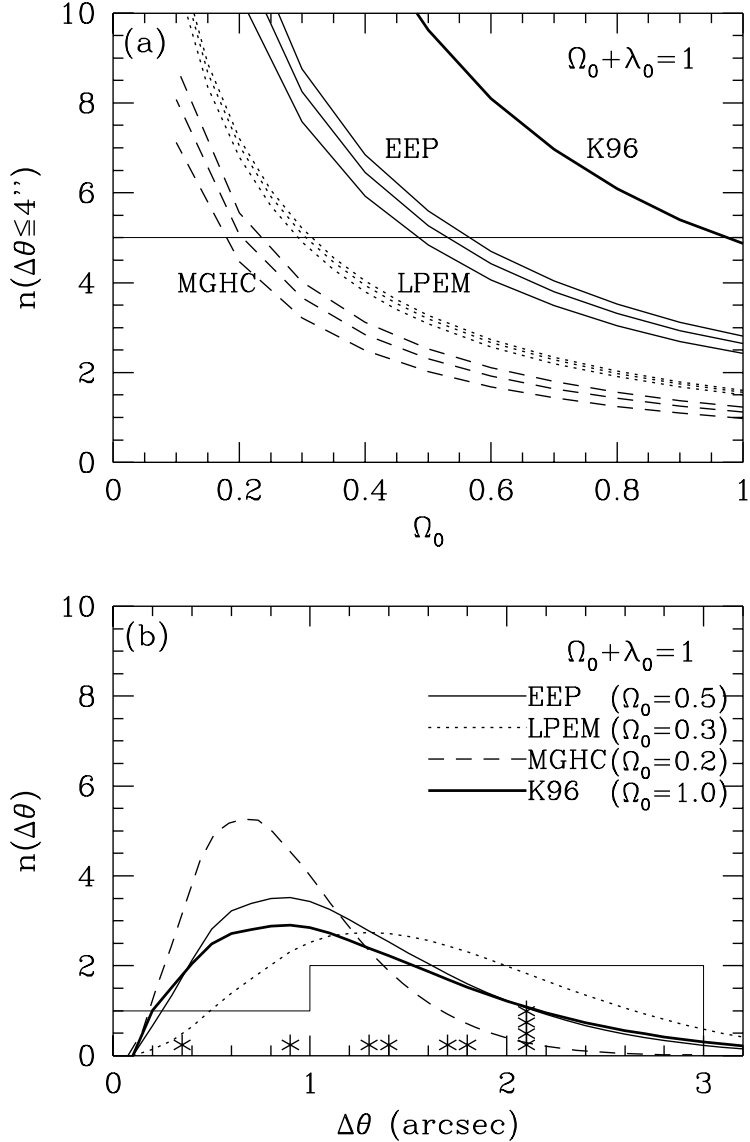}
\caption{
%\figcaption[fig6.ps]{
(a) Predicted total number of lenses $n$ with
$\Delta\theta \le 4''$ in the adopted optical lens surveys, compared with
the observed five lenses (thin solid line). Model calculations are made
for a flat universe $\Omega_0+\lambda_0=1$, and other parameters are given
in the text. Different model curves correspond to the results from the LFs
of EEP (solid lines), LPEM (dotted lines), and MGHC (dashed lines),
with cutoff magnitudes $M^{cut}=-16.5$ (upper lines), $-17$ (middle lines),
and $-17.5$ mag (lower lines), respectively. For comparison, the case for
K96's LF without employing cutoff (thick solid line) is also shown.
(b) Predicted image-separation distribution $n(\Delta\theta)$, compared
with the observed image-separation distribution in the optical sample
(histogram) and in the optical lenses (asterisks located at their
respective separations $\Delta\theta$). The value of $\Omega_0$ for each
model with cutoff $M^{cut}=-17$ mag (except for K96) is chosen so as to
reproduce approximately the observed number of optical lenses, examined
from panel (a).  
}
\end{figure}

\clearpage
\begin{figure}
\plotone{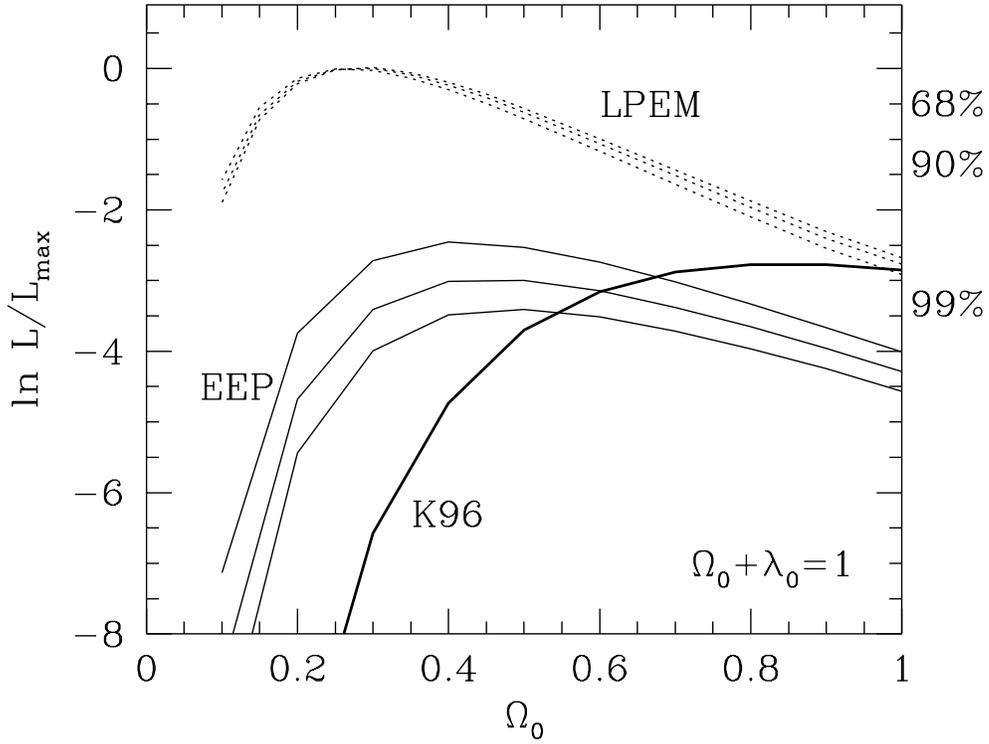}
\caption{
%\figcaption[fig7.ps]{
Results of the maximum likelihood analyses for
reproducing both the total number of optical lenses $n$ with
$\Delta\theta \le 4''$ {\it and} the image-separation distribution
$n(\Delta\theta)$ of optical and radio lenses.
The likelihood functions $L$ defined in eq.(11) are shown as a function
of $\Omega_0$ for a flat universe $\Omega_0+\lambda_0=1$. Different curves
correspond to the results from the LFs of EEP (solid lines), LPEM (dotted
lines), and K96 (thick solid line). For EEP's and LPEM's LFs, three cases
of $M^{cut}=-16.5$, $-17$, and $-17.5$ mag are shown by lower, middle,
and upper lines at the value of $\Omega_0=0.1$, respectively. All likelihood
values are normalized by its maximum $L_{max}$ which is derived using
LPEM's LF with $M^{cut}=-17$ mag. The confidence levels are indicated on the
right-hand side of the plot.
Note that LPEM's LF at $\Omega_0 \simeq 0.3$ yields by far the large
likelihood compared with other cases.
}
\end{figure}

\clearpage
\begin{figure}
\plotone{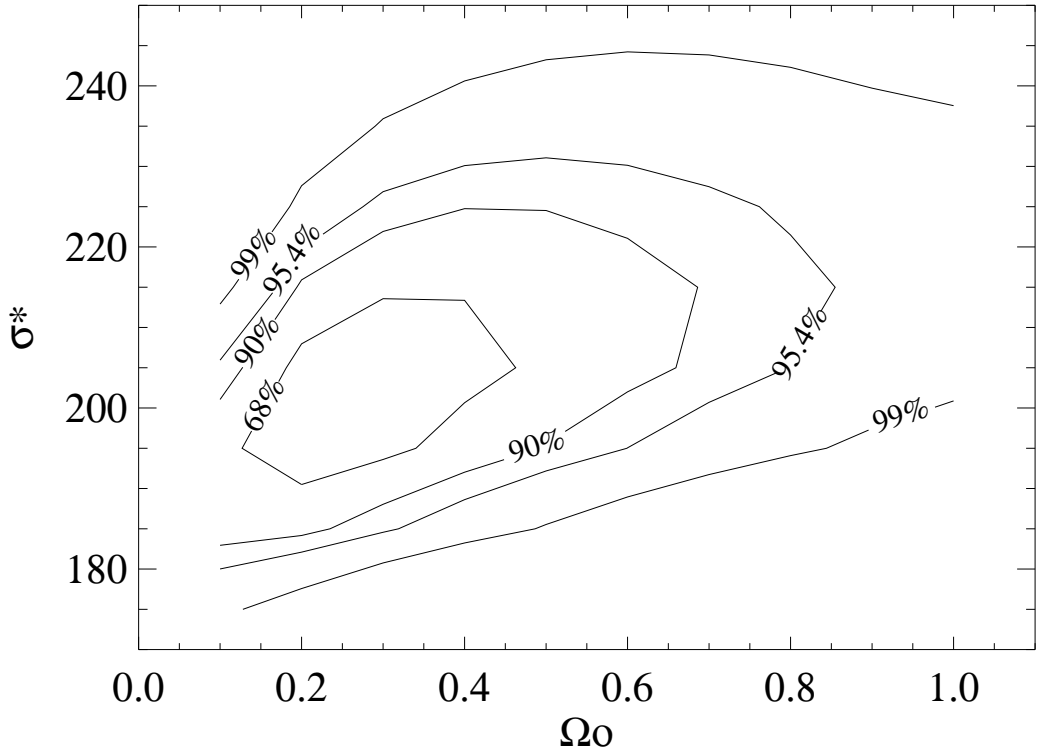}
\caption{
%\figcaption[fig8.ps]{
Likelihood contour plots for flat cosmologies
in the two dimensional parameter space $(\sp^\ast, \Omega_0)$, for our
standard model using LPEM's LF with $M^{cut}=-17$ mag. Contours are shown
at 68 \% (1 $\sigma$), 90 \%, 95.4 \% (2 $\sigma$), and 99 \% confidence
levels for one degree of freedom in the likelihood ratio.
}
\end{figure}

\clearpage
\begin{figure}
\plotone{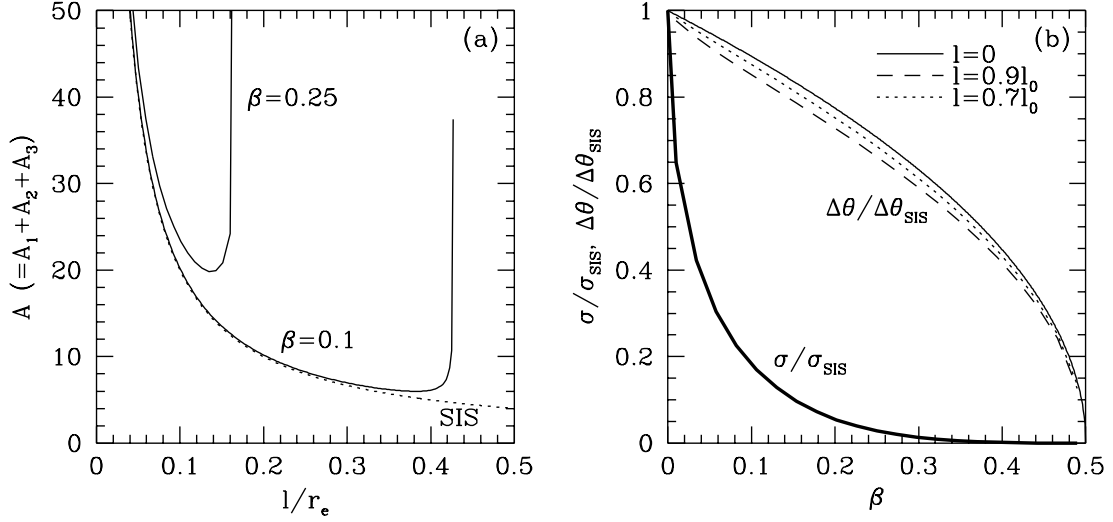}
\caption{
%\figcaption[fig9.ps]{
(a) Total amplification factor $A\equiv A_1+A_2+A_3$
of three lensed images as a function of the source position $l$ normalized
by $r_e$. Solid lines correspond to the cases for non-zero core radii,
$\beta\equiv\rc/r_e=0.25$ and $0.1$, while dotted line for the SIS case
shows the amplification factor of two lensed images.
(b) Ratio of the lensing cross section $\sigma=\pi l_0^2$ to that for the
SIS $\sigma_{SIS}=\pi r_e^2$ (thick solid line) and the ratio of the image
separation $\Delta\theta$ to that for the SIS $\Delta\theta_{SIS}$ (solid,
dotted, and dashed lines), as a function of $\beta$.
}
\end{figure}

\end{document}